\documentclass[twocolumn, secnumarabic,amssymb, nobibnotes, prl, aps,showkeys]{revtex4-2}

\usepackage{graphicx}
\usepackage{amsmath}
\usepackage{float}
\usepackage{lipsum}
\usepackage[normalem]{ulem}
\usepackage{xcolor}

\usepackage{lineno,hyperref}
\modulolinenumbers[5]

\begin{document}

\title{Advancing Time-Resolved Spectroscopies with Custom Scanning Units and Event-Based Electron Detection}

\author{Yves Auad$^{1}$}
\email{yves.maia-auad@universite-paris-saclay.fr}
\author{Florian Castioni$^{1}$}
\author{Jassem Baaboura$^{1}$}
\author{Malo Bézard$^{1}$}
\author{Jean-Denis Blazit$^{1}$}
\author{Xiaoyan Li$^{1}$}
\author{Adrien Teutrie$^{2}$}
\author{Michael Walls$^{1}$}
\author{Odile Stéphan$^{1}$}
\author{Luiz H. G. Tizei$^{1}$}
\author{Francisco de La Peña$^{2}$}
\author{Mathieu Kociak$^{1}$}

\affiliation{$^1$Laboratoire des Physique des Solides, Université Paris Saclay, Orsay, France}
\affiliation{$^2$Unité Matériaux et Transformations, Université de Lille, CNRS, France}

\date{\today}


\begin{abstract}
Direct electron detection is revolutionizing electron microscopy by offering lower noise, reduced point-spread function, and increased quantum efficiency. Among these advancements, the Timepix3 hybrid-pixel direct electron detector stands out for its unique ability to output temporal information about individual hits within its pixel array. Its event-based architecture enables data-driven detection, where individual events are immediately read out from the chip. While recent studies have demonstrated the potential of event-based detectors in various applications in the context of continuous-gun scanning transmission electron microscopes (STEM), the use of such detectors with standard scanning units remains underdeveloped. In this work, we present a custom-designed, Timepix3-compatible scanning unit specifically developed to leverage the advantages of event-based detection. We explore its performance in enabling spatially and temporally resolved experiments, as well as its seamless interfacing with other time-resolved instruments, such as pulsed lasers and electron beam blankers. Additionally, we examine the prospects for achieving enhanced temporal resolution in time-resolved experiments using continuous-gun electron microscopes, identifying key challenges and proposing solutions to improve performance. This combination of custom hardware with advanced detection technology promises to expand the capabilities of electron microscopy in both fundamental research and practical applications.
\end{abstract}


\maketitle


\section*{Introduction}

Since its conception in the 1930s, scanning transmission electron microscopy (STEM) has undergone remarkable advancements in instrumentation, driving the field's success with significant improvements in related techniques such as energy-dispersive X-Ray spectroscopy (EDX), electron energy-loss spectroscopy (EELS), and cathodoluminescence (CL) \cite{kociak2017cathodoluminescence}. Breakthrough innovations, including aberration correction \cite{hawkes2009aberration}, cryogenic electron microscopy (cryo-EM) \cite{murata2018cryo}, and the advent of modern electron monochromators \cite{krivanek2009high, borrnert2023novel}, expanded even further the limits of spatial and spectral resolution as well as the analytical capabilities. These developments have established STEM as an essential tool across materials science and life sciences.

The arrival of direct electron detectors has played an important role in these advancements, offering a reduced point-spread-function and near-unity quantum efficiency in electron detection. One way to categorize these detectors is by how they discriminate individual electron hits and the type of information stored during detection \cite{ballabriga2018asic}. Many utilize a charge-sensitive amplifier, or CSA, which converts deposited charge into a voltage signal. A pixel-by-pixel voltage value, also called threshold, is then compared to the analog levels, ensuring that only signals exceeding this predefined reference are recorded. This significantly increases the signal-to-noise ratio (SNR) by removing spurious background signal. 

The first Timepix \cite{llopart2007timepix} detector was a CMOS-based device that introduced a groundbreaking innovation by reinterpreting the role of the pixel electronics. Instead of simply counting the total number of electron hits, it tracked the number of clock ticks of the internal reference up until the analog levels exceeded the threshold (i.e., an electron hit), recording the so-called electron time-of-arrival (ToA). This enabled precise time-stamping of events, marking a significant departure from conventional designs. Building upon this architecture, the Timepix series has continued to evolve, culminating in the Timepix3 detector \cite{stoffle2015timepix}. With its event-driven design and temporal resolution of $\sim$ 1.56 nanoseconds \cite{poikela2014timepix3, auad2024time}, the Timepix3 opens up a complete new set of possible experiments in electron microscopes.

\begin{figure}[t]
    \centering
    \includegraphics[width=0.45\textwidth]{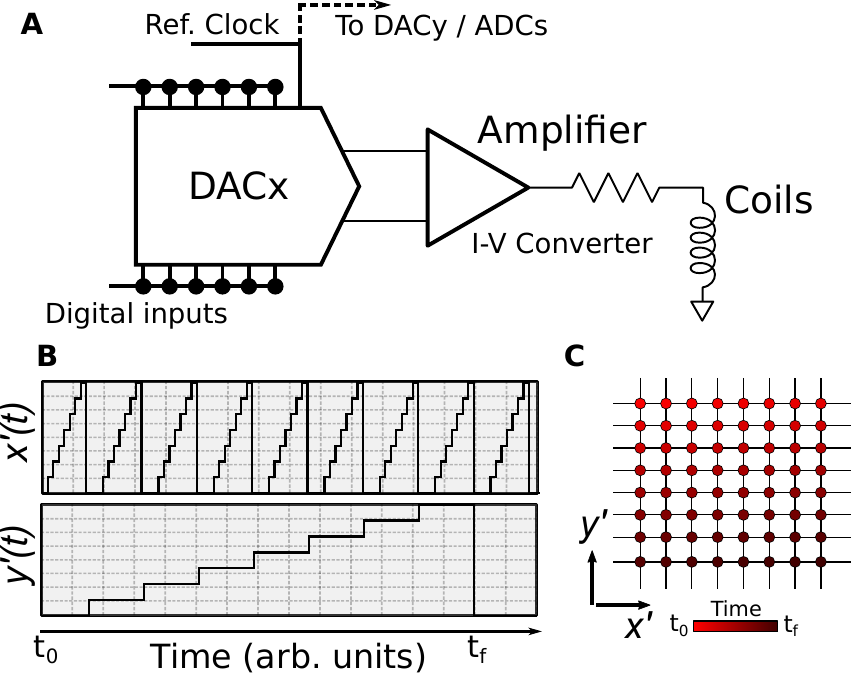}
    \caption{\textbf{The time-dependent scanning unit in scanning transmission electron microscopes.} (\textbf{A}) Scheme of the synchronous DACx driving the microscope coils using an amplifier circuit. The digital inputs come from a digital logic circuit, such as an FPGA. (\textbf{B}) The time-dependent waveforms of a normal rastering scan for both independent scanning directions. (\textbf{C}) Data is shown to the user as function of the electron position $(x', y')$. Although time-dependency is not explicit, both $y'(t)$ and $x'(t)$ are fundamental waveforms defined by the scanning pattern.}
    \label{Figure0}
\end{figure}

The Timepix3 detector is particularly well-suited for integration with scanning electron probe systems, such as STEMs. Unlike conventional transmission electron microscopes, STEM relies on spatially-determined, focused electron probes, the movement being achieved through the use of scanning units (SU), as shown in the simplified functional scheme in Figure \ref{Figure0}. SUs use digital-to-analog (DAC) and analog-to-digital (ADC) converters, which are synchronous logic circuits in which the data conversion is latched to a reference clock, typically running between 25 and 200 MHz, thereby conferring tens of nanosecond temporal resolution on such devices (Figure \ref{Figure0}A). As an example, the standard top-to-bottom, left-to-right raster scan is done by incrementing the appropriate values to the DACs for both the X and Y coordinates. On the next clock tick, the DAC's output drives the microscope coils through the amplifier circuit. In essence, scanning units create time-dependent waveforms, $x'(t)$ and $y'(t)$, that define the spatial behavior of the electron probe, as shown in Figure \ref{Figure0}B. Typical data acquisition from an imaging device such as an annular dark-field (ADF) detector is done by an ADC latched on the same reference clock. Because all ADCs and DACs are synchronous and controlled by the same reference, it is possible to reconstruct an image based on the electron probe position $(x', y')$, as shown in the panel C of Figure \ref{Figure0}, abstracting time dependency. By recognizing the inherently time-based nature of scanning units, it is clear that interfacing them with a time-tagging detector like Timepix3 is a logical choice. The Timepix3’s capability to record precisely the time of electron arrivals aligns naturally with the time-dependent position of the electron probe.

\begin{figure}[b!]
    \centering
    \includegraphics[width=0.45\textwidth]{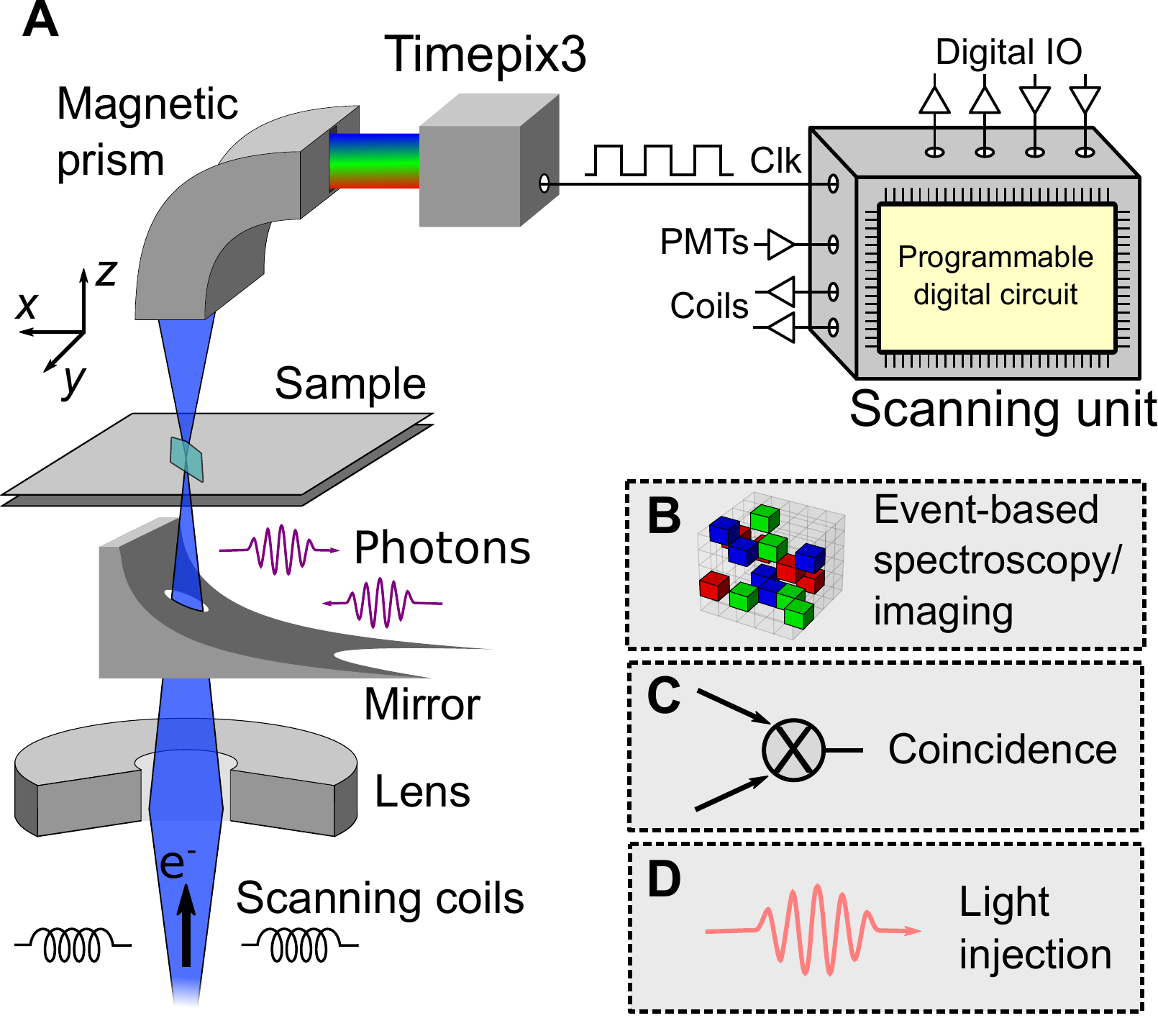}
    \caption{\textbf{Scheme of the Timepix3 and the scanning unit interface.} (\textbf{A}) Schematic of the STEM is shown, including the scanning unit, an EELS spectrometer followed by a Timepix3, and a parabolic mirror for light detection (CL) and injection (EEGS). The scanning unit is responsible for driving the electron probe position of the STEM, as well as acquiring analog signals, typically from a photo-multiplier tube. By putting Timepix3 in the same clock domain, the time-of-arrival of the electron hits in the detector can be related to the current position of the focused electron probe. Several experiments can be performed using this Timepix3 - SU architecture, including (\textbf{B}) fast hyperspectroscopy or 4D-imaging using arbitrary but well-defined time-dependent waveforms in each one of the microscope scanning coils, (\textbf{C}) Electron-photon coincidence measurements, by either inputting an external sample stimulus, or by receiving it, e.g. CL photons and EELS electrons, and (\textbf{D}) time-resolved light-injection experiments, such as in nanothermometry or in electron energy-gain spectroscopy.}
    \label{Figure1}
\end{figure}

In this work, we explore the wide range of experiments enabled by interfacing TPX3 with our custom-designed scanning unit, as illustrated in Figure \ref{Figure1}. A key capability is the adaptation of Timepix3 to arbitrary scanning patterns, including non-regular spatial sampling of the region of interest (ROI), which is presented first. The combination of event-based detection and dedicated scanning units enables flexible event-based spectroscopy and imaging. This is illustrated for different new types of spectroscopies allying EELS and photon detection or injection. A prime example is cathodoluminescence excitation spectroscopy (CLE) \cite{varkentina2022cathodoluminescence} that utilizes time-coincidence measurements to correlate electron arrivals with emitted photons, providing a powerful tool for studying electron-photon interactions. This is also the case for electron energy-gain spectroscopy (EEGS) \cite{de2008electron}, which is enabled by synchronized  light injection experiments. Finally, we will present recent applications in spatially- and time- resolved thermometry that offer insights into local temperature distributions.

\section*{Timepix3 and the scanning unit integration}

Commercially available SUs typically feature at least one input logic pin, enabling frame-based synchronized experiments with the installed electron detector. This approach is commonly used in techniques such as hyperspectral imaging and 4D-STEM. In traditional frame-based experiments, the detector acts as the leader, triggering the probe position to advance to the next pixel after completing a frame acquisition, with the scanning unit controlling the electron probe as the follower. However, the Timepix3 detector introduces a paradigm shift with its superior temporal resolution, warranting a reversal of this leader/follower relationship. In the new configuration, the scanning unit becomes the leader, defining the electron probe's position in the time domain, while the Timepix3 acts as the follower, precisely capturing the electron ToA and mapping it to the correct $(x', y')$ position. Any electronic delay between the SU and the detector can be accounted for as a constant offset applied to the measured ToA.

The optimal approach to integrating it as an SU-follower depends on the specific manufacturer's implementation. A widely used readout system, SPIDR3 \cite{van2017spidr}, includes two supplementary lines of time-to-digital converters (TDCs), enabling external events to be timestamped using the same reference clock as the one applied for electrons. When the SU operates as the leader, periodic pulses from the SU can be used to determine the electron probe position corresponding to a given Timepix3 clock tick, as detailed in prior studies \cite{auad2022event}. While the SU and Timepix3 do not strictly share a common base clock, the resulting synchronization effectively achieves identical outcomes.


\subsection*{Development \& Characteristics of a Timepix3-era Scanning Unit}

The primary application of event-based detectors is to exploit their impressive temporal resolution alongside the intrinsic temporal resolution of the SU. With synchronized clocks, performing spatially resolved electron detection is trivial, which has already been shown in previous works, for both hyperspectral EELS imaging \cite{auad2022event} and for 4D-STEM datasets \cite{jannis2022event}.

The Timepix3 detector, being faster than the scanning unit, has the potential to push the performance of the SU to its limits. However, traditional challenges associated with scanning patterns must still be addressed to fully exploit the detector's capabilities. Various scanning patterns and their related challenges and solutions have been extensively studied in the literature, including serpentine rastering patterns \cite{ortega2021high}, random scans \cite{zobelli2020spatial}, spiral scans \cite{sang2017precision}, and Lissajous-based scanning \cite{li2018compressed}.

\begin{figure}[t]
    \centering
    \includegraphics[width=0.45\textwidth]{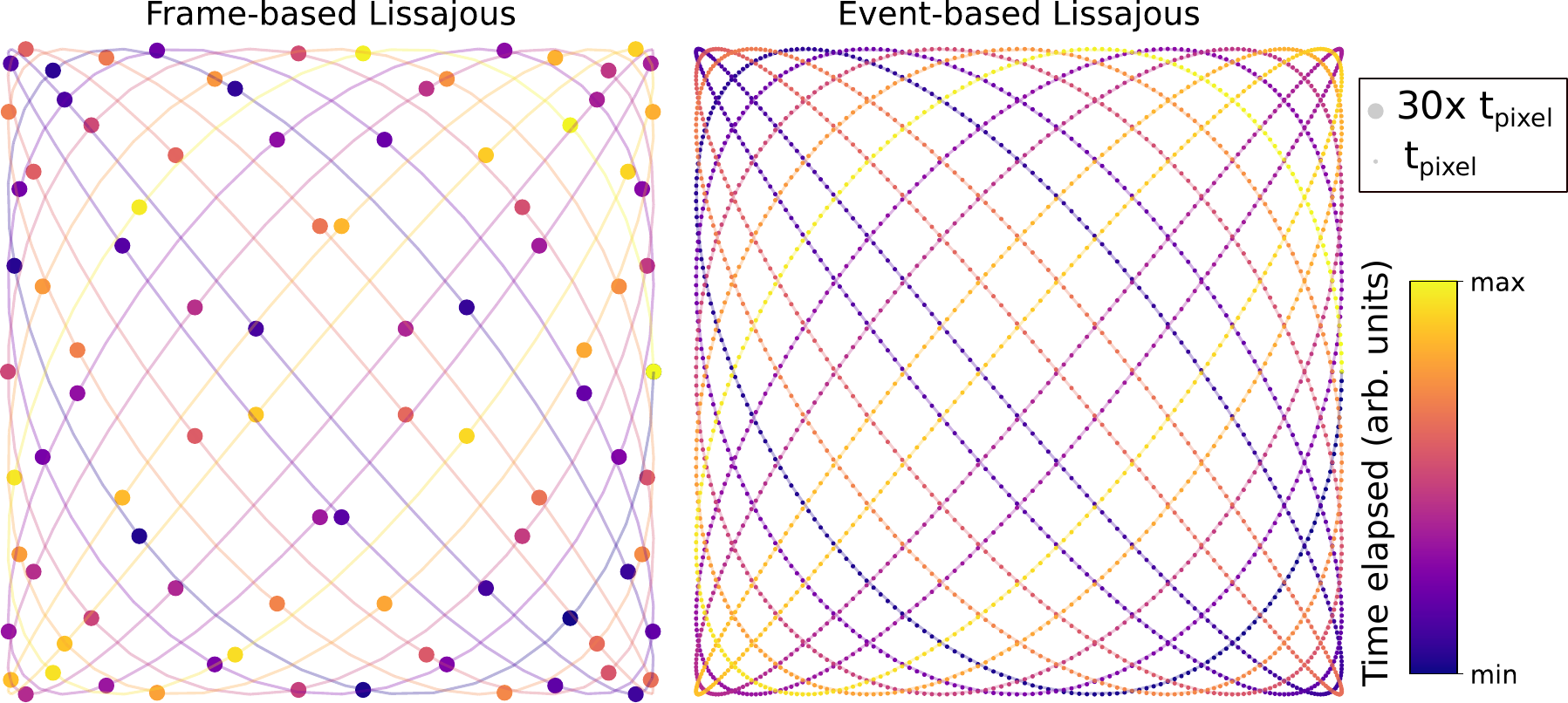}
    \caption{\textbf{Comparison between frame and event-based in Lissajous scanning for a fixed total frame time.} Lissajous scanning can be used with frame-based detectors, but the minimum attainable detector dwell time limits the highest achievable Lissajous frequency. When a nanosecond-resolved electron detector is used, higher frequencies become possible. In the frame-based case, the electron position is fixed until the detector frame acquisition is finished, and the pixel dwell time is 30 times longer than in the event-based case, also indicated by the radius of the scatter sphere.}
    \label{FigureLissajousComparison}
\end{figure}

Alternative scan patterns are of great interest due to their potential to significantly influence the dissipation of accumulated deposited dose \cite{egerton2004radiation, jones2018managing, jannis2022reducing}. Sparse sampling has been suggested to reduce beam damage, and together with compressed sensing and inpainting techniques, could provide significant improvements in this direction. One study, using random scan patterns, has been shown to improve the irradiation damage in the cathodoluminescence signal of an hexagonal boron-nitride (\textit{h}--BN) flake \cite{zobelli2020spatial}. However, due to their broad frequency-domain response, the scanning system suffers from a limited pixel dwell time. On the other hand, sinusoidal-based systems, such as Lissajous scanning, offer great opportunities in electron microscopy. The proper implementation of such systems can be challenging, with the need for fast digital reference clocks, and high data throughput logical circuits. 

Figure \ref{FigureLissajousComparison} compares Lissajous scanning using either frame-based or event-based detection. The total frame acquisition time is fixed across all cases. Although Lissajous scanning can be employed with frame-based detectors, this approach is constrained by the two triggering configurations. If the camera triggers the scanning unit, the long dwell time required by the detector limits the highest achievable frequency (Figure \ref{FigureLissajousComparison}, left panel). If the scanning unit triggers the camera, the beam's movement during frame acquisition transforms discrete probe points in the left panel of Figure \ref{FigureLissajousComparison} into line segments, which results in a blurred image. In contrast, event-based detectors—such as Timepix3—do not have this limitation and can achieve significantly higher frequencies without compromising the image resolution. The theoretical maximum frequency is determined by the detector’s temporal resolution ($\sim 1.56$ ns), but in practice, it is limited to a few hundreds of kHz because of the inductance of the scanning coils and the bandwidth of the electronics. There are other options to design scanning systems, such as electric deflectors. This system would equally produce a low-pass filter behavior and the maximum attained frequency will be equally limited by the electronics bandwidth but also by the deflector capacitance.  



\begin{figure}[t!]
    \centering
    \includegraphics[width=0.45\textwidth]{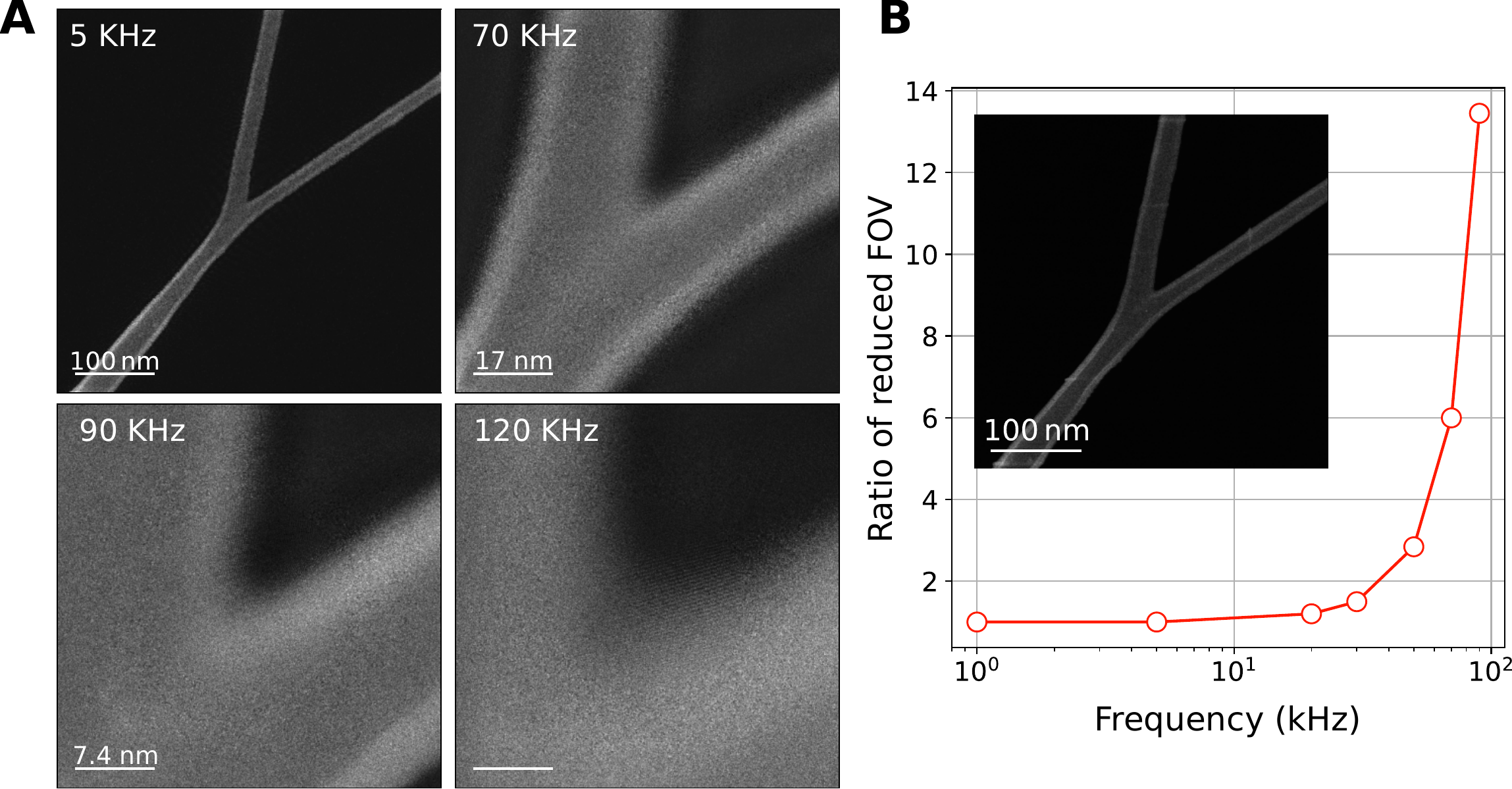}
    \caption{\textbf{Lissajous-based scanning patterns.} (\textbf{A}) A series of Lissajous scans ($512 \times 512$ pixels) and their respective sinusoidal frequencies, from 5 kHz up to 120 kHz, displayed in real time to the user. Despite some degradation of the image, its features remain visible. However, at 120 kHz, the loss of reference prevented proper calibration. (\textbf{B}) Due to the circuit bandwidth, coil impedance rises enough to reduce the current through the coil, reducing the field of view, as can be seen by the curve. In the inset, the image of the region of interested obtained by normal rastering pattern.}
    \label{FigureLissajous}
\end{figure}



Figure \ref{FigureLissajous} presents a series of images ($512 \times 512$ pixels) of a lacey carbon sample generated by linear interpolation from Lissajous scan patterns at scanning frequencies of 5, 70, 90, and 120 kHz on a Nion Hermes 200 microscope. Because Lissajous sampling is not evenly distributed across the field of view, linear interpolation was performed to produce the images, which are displayed in live conditions to the user. This represents an approximately hundred-fold increase in frequency compared to previous STEM Lissajous scanning experiments. Compared to a standard raster scanning system, as shown in \ref{Figure0}B, the probe speed for Lissajous scanning is significantly faster—approximately 30 times in the X-coil direction and 5 $\times 10^4$ in the Y-coil direction. Since Lissajous scanning applies similar frequency components to both scanning coils, the vertical scan direction (Y axis) achieves an improvement in speed exceeding four orders of magnitude. Moreover, the similar frequency components ensure a comparable amplitude-phase response for both coils, resulting in symmetrical growth in coil impedance in both directions. Consequently, the observed field of view (FOV) decreases symmetrically in both directions, preserving image features despite the reduction in FOV. 

\subsection*{Event-based hyperspectral imaging}

\begin{figure}[t!]
    \centering
    \includegraphics[width=0.45\textwidth]{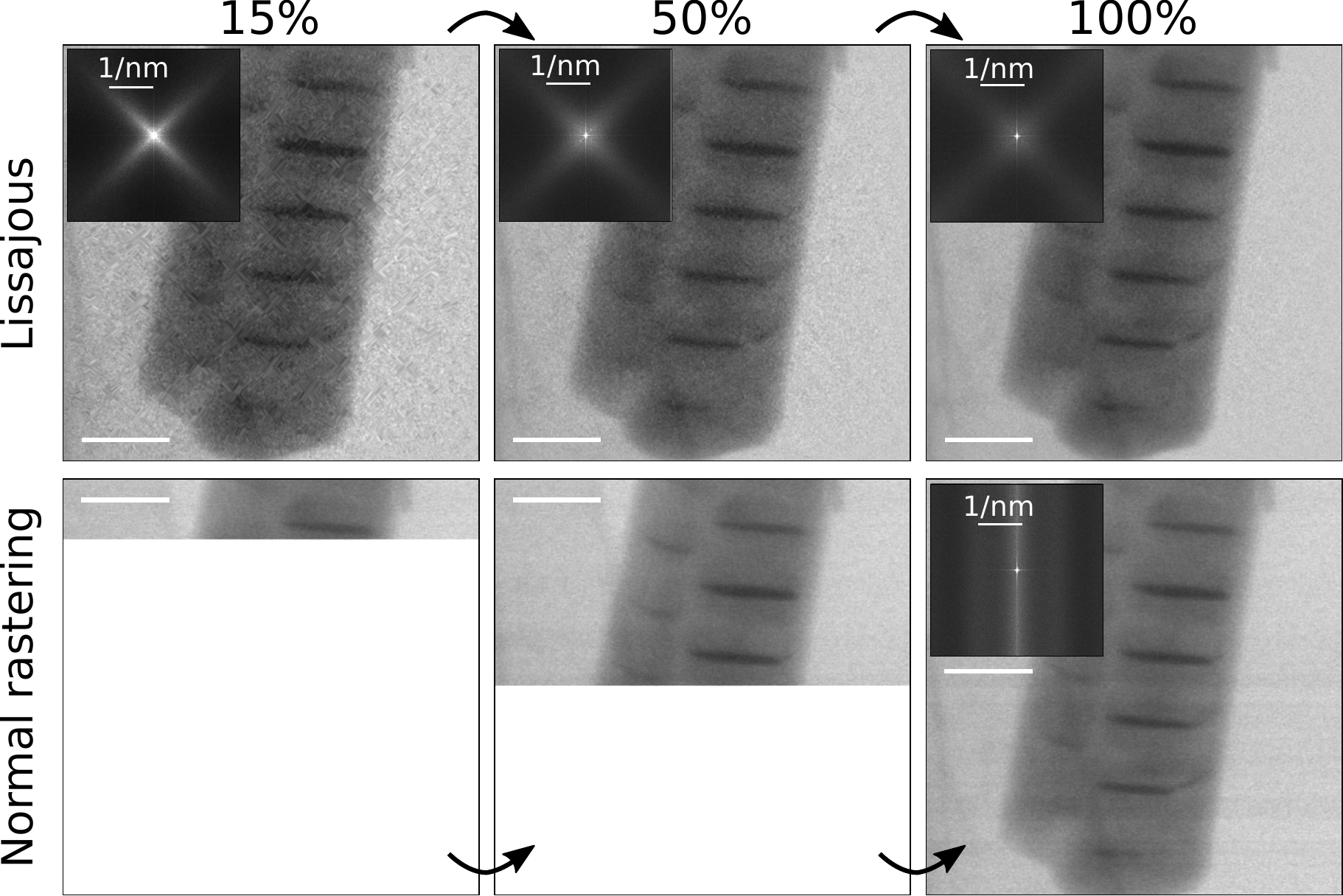}
    \caption{\textbf{Comparison between Lissajous scanning and normal rastering.} Lissajous scanning enables rapid, multi-resolution imaging. As the scan progresses, the spatial resolution improves in Lissajous scanning by collecting more sample points. The imaging process transitions from sparse to dense sampling, allowing traditional sparse-sampling techniques, such as inpainting and interpolation, to be applied during the initial phase of image acquisition. In both scanning patterns, the total deposited dose is the same, but because of the flyback time, normal rastering is $\sim$ 30\% slower. Lissajous scanning operates at approximately 8 kHz. Scale bars are 100 nm.}
    \label{FigureLissajousMulti}
\end{figure}

\begin{figure*}[t!]
    \centering
    \includegraphics[width=0.55\textwidth]{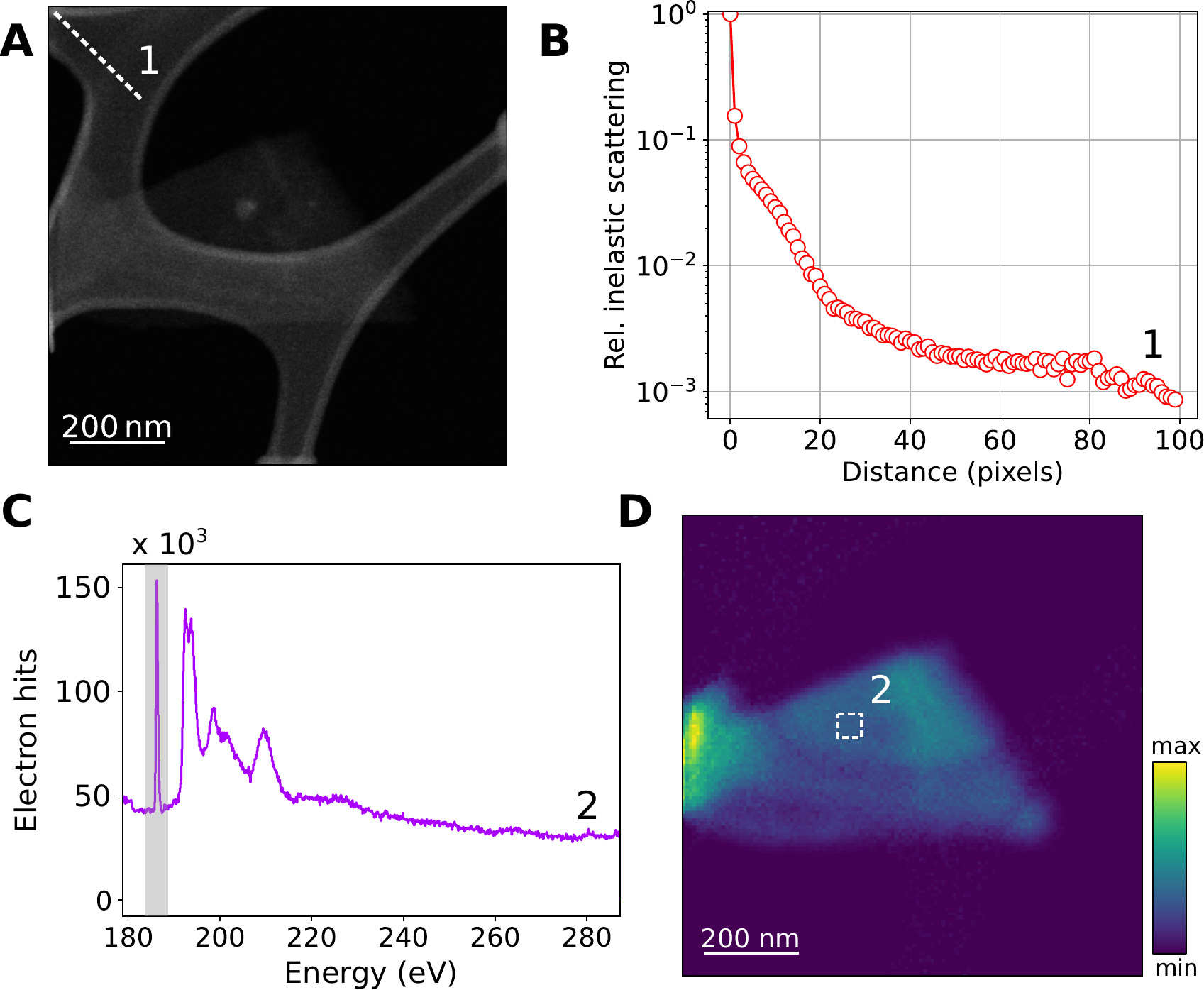}
    \caption{\textbf{Hyperspectral imaging with Lissajous scanning and the Timepix3 event-based detector, displayed live to the user.} (\textbf{A}) The ADF image of the region of interest. (\textbf{B}) The total number of inelastically scattered electrons between 180 and 290 eV in the profile trace in (\textbf{A}) marked 1. Lissajous scanning is spatially inhomogeneous, and the deposited dose is higher in the corners than in the center of the image. (\textbf{C}) The associated spectrum from  the region of interest marked as 2 in (\textbf{D}). (\textbf{D}) The hyperspectral image integrated over the energy window between 185 and 190 eV, comprising the $\pi$* signature of \textit{h}--BN.}
    \label{FigureLissajousSPIM}
\end{figure*}

One of the most compelling aspects of Lissajous scanning is its ability to enable rapid imaging, allowing microscopy images to be reconstructed before all scanning points are acquired. These systems are often described as achieving multiresolution imaging, as the spatial resolution depends on the number of points considered \cite{tuma2012high}. In practice, Lissajous scanning acts as a sparse sampler in the early stages of image acquisition and gradually transitions to denser sampling. Traditional sparse-sampling image reconstruction techniques, such as signal inpainting or interpolation, can be applied throughout the data acquisition process. Figure \ref{FigureLissajousMulti} illustrates this concept by comparing images taken from Lissajous and conventional raster scanning on GaN nanowires with AlN quantum disks, acquired in a vacuum generators (VG) HB501 microscope. In both cases ($2048 \times 2048$ points, $250$ ns per pixel), the total deposited dose is the same, but because of the flyback time of the normal rastering, frame acquisition time is smaller in Lissajous scanning ($\sim$ 1 s for the latter, and $\sim$ 1.3 s for the former). Finally, the percentages shown at the top of the image refer to the relative dose deposition. With only 15\% of the total dose, the overall structure of the GaN/AlN nanowire can be reconstructed using Lissajous scanning, and the spatial resolution is already sufficient to distinguish the AlN disks. As the scan progresses and additional data points are acquired, spatial resolution improves, and the image becomes clear, as is also shown by the inset absolute values of the fast-fourier transform images. In this particular example, images were reconstructed using linear interpolation, but more advanced reconstruction techniques could potentially yield better results \cite{li2018compressed, ede2020partial}. The time required for interpolation for $2048 \times 2048$ pixels is comparable to that of the entire image acquisition and, for this reason, Figure \ref{FigureLissajousMulti} was generated through post-processing. Nonetheless, opportunities exist for software acceleration that would enable partial image reconstructions to be displayed live during acquisition, as it is the case for smaller images such as the ones displayed in Figure \ref{FigureLissajous}. It is also interesting to note that normal rastering seems to provide a better contrast than Lissajous scanning. Although the total deposited dose and the number of sampled points is equal, Lissajous scanning has a non-homogeneous sampling, meaning that resolution is also spatially dependent. In the Lissajous scan of Figure \ref{FigureLissajousMulti}, $99.9\%$ of the consecutive probe position are separated by more than 1.5 \AA, already larger than the expected probe size in this microscope of roughly 1 \AA. The average inter-point distance is about 30 \AA, with a standard deviation of 8.5 \AA. For the raster scan, consecutive pixel distances along the line is 2.4 \AA. This illustrates a key feature of the Lissajous approach, as its average inter-point distance is larger relative to raster scanning.

A comparison can be drawn between the Lissajous scanning and the event-driven nature of the Timepix3 detector. In imaging terms, each newly acquired pixel contributes with unique information to the image reconstruction. Consequently, depending on the required spatial resolution, Lissajous-based rastering achieves exceptional temporal resolution, often within subframe timescales. This approach is particularly advantageous for dose-sensitive samples, as data can be collected only up to a predefined damage threshold, with the temporal resolution determined by the pixel dwell time. Moreover, the enhanced temporal resolution also enables sub-frame drift correction of the electron probe.

With this synergy in mind, we have studied the feasibility of this approach, synchronizing Timepix3 and the newly developed SU, as shown in Figure \ref{FigureLissajousSPIM} in the case of hyperspectral EELS imaging of an \textit{h}--BN flake, displayed to the user in live conditions. The list of points defined by a Lissajous-based scan (running at $\sim$ 30 kHz) is sent to Timepix3 at the beginning of the acquisition. Depending on the electron time-of-arrival, the electron is unequivocally assigned to an electron probe position. In panel \ref{FigureLissajousSPIM}D, the image contrast comes from the Gaussian fitting amplitude of the $\pi$* peak in the K-edge of \textit{h}--BN.

The ADF image of the region of interest is shown in Figure \ref{FigureLissajousSPIM}A, and the highlighted traced line 1 shows the spatial profile along it of the total number of inelastic electrons between 180 and 290 eV, shown in panel \ref{FigureLissajousSPIM}B. Although not a direct measure of the total electron-deposited dose, this profile along a roughly homogeneous thickness shows one of the most undesired aspects of the Lissajous-based scan, i.e. the uneven irradiation of the region of interest. In contrast with a standard rastering, where the scanning speed is roughly constant in both scan directions, sinusoidal curves in Lissajous scanning have time-dependent acceleration waveforms. In the crest of the waveforms (borders of the image), the probe decelerates, sampling many neighbor points in this region in a given time, and this effect is particularly strong when both scanning axes are at their crests, which occurs at the corners of the image area. It is important to note that the dose inhomogeneity is not directly apparent in the ADF images in Figures \ref{FigureLissajous}/\ref{FigureLissajousSPIM}, as these images are reconstructed through linear interpolation.

In panel \ref{FigureLissajousSPIM}D, we present the hyperspectral imaging results associated with the region of interest defined by the grayed energy range in panel \ref{FigureLissajousSPIM}C, which corresponds to the $\pi$* excitation signature of the K-edge of \textit{h}–BN. The data is not normalized for the inhomogeneous dose deposition of Lissajous scanning, leading to contrast skewed towards the sample edges. In this study, we have not conducted an in-depth investigation into whether these newly developed fast scanning patterns effectively mitigate sample damage. However, we plan to address this aspect in future work. 

It is relevant to note that dose normalization with an event-based detector can be achieved by counting the number of electrons per pixel. While some pixels may experience damage earlier than others, having a complete time-resolved record for each pixel allows us to account for these variations. Additionally, in Lissajous scanning, the trajectory is approximately linear in the central region, and the beam can be blanked when scanning the outer parts of the sample, thereby achieving a more homogeneous dose deposition.

Having validated the principle of a using a time-resolved EELS detector in conjunction with a dedicated SU for ns time-resolution imaging and EELS, we switch to other spectroscopies made possible by adding coincidence or synchronization of electrons with photons.

\section*{Electron-photon coincident pairs}

A common technique in optical spectroscopy is photoluminescence excitation spectroscopy (PLE) \cite{hill2015observation}. It uses a tunable laser to excite a sample while the emission spectra are recorded. This way, the path from the creation of an optical excitation by absorption at a given energy to its emission at a different one can be recorded. Beyond the physical insights it offers, PLE is a technique of choice for various applications, e.g. measuring the quantum efficiency of materials. PLE, however, remains diffraction-limited, and thus cannot address nanostructured materials. 

Because the fast electron provides a broadband energy excitation, from far-infrared molecular vibrations ($<$100 meV)  \cite{krivanek2014vibrational} to over 10 keV \cite{maclaren2018eels}, one can perform the equivalent in electron microscopy (with EELS electrons signaling energy absorption, and CL photons energy re-emission). However, this technique, which we call cathodoluminescence excitation spectroscopy (CLE), is not straightforward. One way to achieve it is by temporally correlating EELS electron and cathodoluminescence photon pairs \cite{varkentina2022cathodoluminescence, feist2022cavity, varkentina2023excitation} (as shown in Figure \ref{Figure1}B), provided that the electron energy loss is known. This can be accomplished using the Timepix3 detector and a single-photon-counting photomultiplier tube for precise time correlation. Additionally, CLE benefits from much of the standard experimental setup used in cathodoluminescence inside an electron microscope, simplifying its implementation.

\begin{figure}[t]
    \centering
    \includegraphics[width=0.45\textwidth]{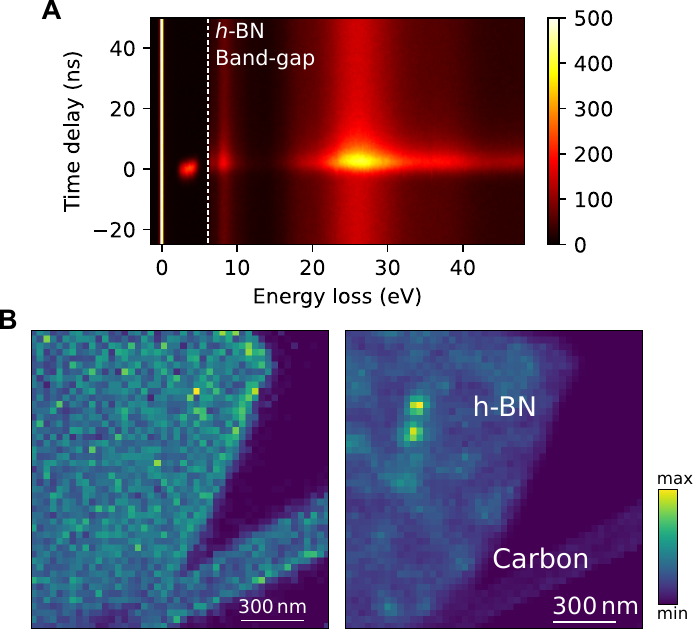}
    \caption{\textbf{Temporally coincident inelastic electrons and emitted photons provide access to new spectroscopic information.} (\textbf{A}) Time delay between the emitted (CL) photon and the EELS electron as a function of the energy loss, with the \textit{h}--BN band-gap identified by the white dashed line. (\textbf{B}) CLE filtered images. The images show an \textit{h}--BN sample on an amorphous carbon layer using an energy window below (left) and above (right) the \textit{h}--BN bandgap. For electrons with insufficient energy, below the band-gap, the photon emission process is mostly due to transition radiation with minor contrast variations and slight delocalization. When the electron energy is above the gap, \textit{h}--BN intrinsic defect emission plays a role, as highlighted by the two bright spots. Panel B is reproduced with permission from ref. \citenum{varkentina2022cathodoluminescence}  (Copyright 2022 American Association for the Advancement of Science).}
    \label{Figure_CLE}
\end{figure}

Figure \ref{Figure_CLE} illustrates such an experiment \cite{varkentina2022cathodoluminescence}, in which a thin \textit{h}–BN sample on an amorphous carbon layer was studied. Panel A shows the time delay between the emitted photon and the inelastically scattered electron as a function of the electron energy loss, the \textit{h}--BN band-gap energy ($\sim$ 6 eV) is highlighted by the vertical white dashed line. Below the band-gap, the CLE signal is dominated by an energy peak attributed to transition radiation \cite{varkentina2022cathodoluminescence,scheucher2022discrimination}. At energies higher than the band-gap, the volume plasmon dominates. We note here that the typical lifetimes of the emitters, which are known to be point defects \cite{Bourrellier}, are smaller than the time resolution of the Timepix3 \cite{auad2024time}. CLE can also be used to measure lifetimes when they are larger than about 2 ns \cite{varkentina2023excitation}. By selecting electrons within a narrow time window ($\pm 5$ ns) around the detection of a photon event, a hyperspectral image of correlated electrons was reconstructed. In the left-hand figure of \ref{Figure_CLE}B, the contrast is determined by electron energy losses below the band-gap of the material. The very weak variations in the signal and slight delocalization confirm that the contrast is dominated by the transition radiation \cite{varkentina2022cathodoluminescence}. On the right is displayed a map filtered at energies higher than the band-gap. In particular, the appearance of  two bright point defects suggests that these higher-energy electron losses are the source of the emissions from defect states intrinsic to the material, as previously predicted \cite{Meuret2015} but never proved experimentally.

Correlating electron-photon pairs can also be interesting in a variety of other experiments. For example, it has been suggested that detecting X-Rays and electron coincidences could significantly suppress the background, on both EELS and EDS spectra \cite{kruit1984detection, jannis2021coincidence}. Additionally, electron-photon pairs have been used for enhanced imaging contrast in photonic cavities \cite{feist2022cavity}, heralding non-classical light \cite{arend2024electrons}, and nanosecond-scale spatially-resolved optical transition lifetime measurements \cite{varkentina2023excitation}. By knowing the relative time between the inelastically scattered electron and the emitted photon, it is possible to perform such experiments. In particular, using Timepix3 allows one to perform this experiment in a multi-channel electron detection fashion, being able to timestamp all incoming electrons.

\section*{Light-injection experiments in a continuous-gun electron microscope}

\begin{figure}[b]
    \centering
    \includegraphics[width=0.45\textwidth]{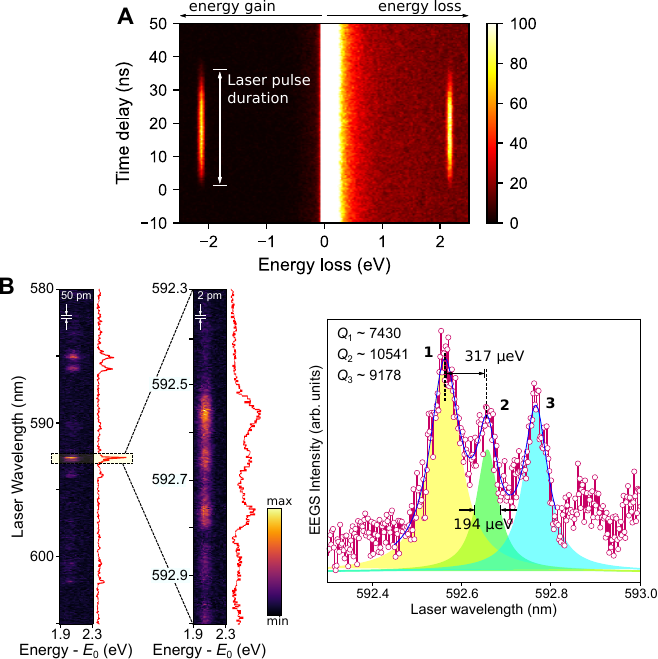}
    \caption{\textbf{By combining PINEM and EEGS one can study sub-meV optical resonances with nanometer spatial resolution.} (\textbf{A}) The time delay between the laser pulse trigger and electron energy loss/gain as a function of the electron energy. In this particular experiment, the optical excitation lifetime is much shorter than the laser pulse ($\sim 30$ ns). (\textbf{B}) Electron energy-gain spectroscopy is performed in a whispering-gallery mode microresonator to unveil a $\sim 100$ $\mu$eV wide resonance. Panel B is reproduced with permission from ref. \citenum{auad2023muev} (Copyright 2023 Springer-Nature).}
    \label{Figure_GAIN}
\end{figure}

The principle of correlating EELS electrons and CL photons can be inverted, by performing synchronized experiments between an injected laser photon and EELS electrons (Figure \ref{Figure1}D). By doing so, one can perform EEGS experiments \cite{de2008electron, asenjo2013plasmon}, which constitute a spectrally resolved version of photon-induced near-field electron microscopy (PINEM) \cite{barwick2009photon}. In the latter, sub-ps laser and electron pulses illuminate a sample of interest in a synchronized way, triggering both stimulated energy gains and losses. If the laser energy $\hbar \omega$ is changed, and the intensity of the stimulated gain or loss peak recorded, one can record a full EEGS spectrum with an energy resolution limited only by the laser spectral resolution. Due to the energy-time Heisenberg uncertainty, where $\Delta t \sim ps$ corresponds to $\Delta E \sim meV$, high energy resolution cannot be reached with regular pulsed gun setups \cite{Pomarico2018,Wang2020}. A solution with a continuous beam and longer pulse width (at least ns) lasers is therefore needed.

Light-injection experiments in a continuous-gun microscope require a method to correlate the detected electrons with the injected laser pulse. Initial experiments on the subject have synchronized the laser trigger and a pair of electrostatic beam blankers placed just before the EELS electron detector \cite{das2019stimulated, auad2023muev}. While this setup required significant modifications to the microscope and a high numerical aperture mirror, it enabled the experiment to be performed independently of the temporal resolution of the electron detector. A digital delay line was employed to control the time shift between the logic signals to the laser and the blanker, with approximately 10 ns precision. Due to the proximity of the deflection system to the detector, the voltages applied to the system had to swing by hundreds of volts in square waveforms, limiting the system's speed to only a few tens of kHz.

Using Timepix3 eliminates the need for a fast blanker, requiring only the identification of the electron time-of-arrival relative to the laser trigger—a straightforward task when the SU and Timepix3 are properly synchronized. This is demonstrated in panel A of Figure \ref{Figure_GAIN}, where a ~30 ns laser pulse is used. The histogram shows the time delay between electron arrival and laser triggering as a function of electron energy loss. The stimulated energy loss and gain are visible at ±2.1 eV, corresponding to the 585 nm laser photon energy. 

\begin{figure}[t]
    \centering
    \includegraphics[width=0.45\textwidth]{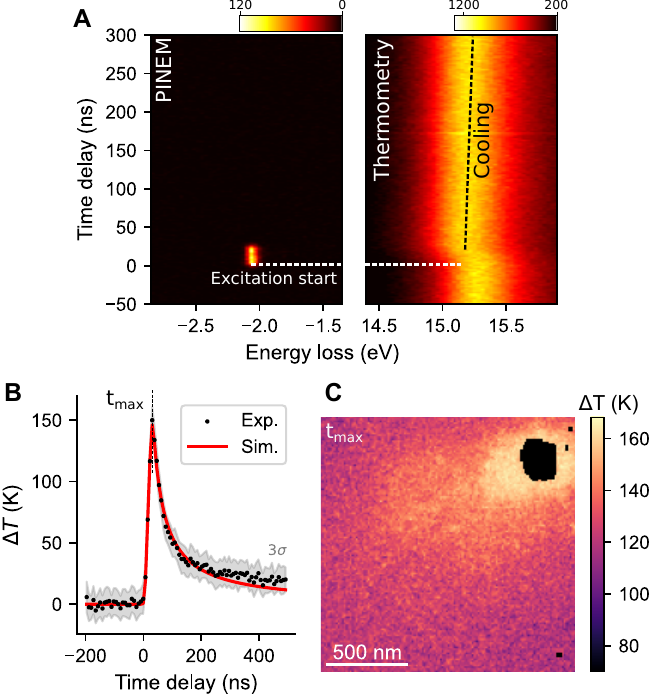}
    \caption{\textbf{Nanoscale-resolved thermometry experiments performed in an electron microscope.} (\textbf{A}) The time delay between the laser pulse trigger and the electron energy loss/gain as a function of the electron energy. In this experiment, performed on an Aluminum film, PINEM peaks are used to identify the laser pulse, and hence the start of the locally induced heating. In the cooling period, the Al bulk plasmon energy slowly shifts back to the equilibrium value. (\textbf{B}) The profile of the sample cooling as a function of the time delay. (\textbf{C}) Synchronized with the custom-developed SU, spatial histograms of temperature can be drawn at a given time delay. Panel B and C is reproduced with permission from ref. \citenum{castioni2025nanosecond} (Copyright 2025 American Chemical Society).}
    \label{Figure_PLASMON}
\end{figure}

Two recent experiments have shown EEGS experiments with spectral resolution in the $\mu eV$ range, surpassing the state-of-art electron monochromators \cite{dellby2020ultra}. In the first, a sample holder has been adapted to fit a high quality-factor ring resonator and a light-pumping system with a waveguide \cite{henke2021integrated}. In the second experiment, a parabolic reflector is used to focus the laser in a small spot at the sample level. The free-space coupling is orders of magnitude less efficient than in the former example, but it allows one to perform such experiments on arbitrary samples, not needing special holders. Additionally, the weaker light coupling is mitigated by the use of an electron monochromator to increase the energy-gain signal-to-background ratio \cite{auad2023muev}. This approach has been applied to whispering-gallery mode microresonators drop-cast on standard TEM lacey carbon grids \cite{auad2022unveiling}. By accurately positioning a parabolic reflector, we have detected optical resonances in the range of $\sim 100$ $\mu eV$, as shown in Figure \ref{Figure_GAIN}B.

We note that, besides PINEM and EEGS, light injection systems remain to be explored in many other applications. They have potential uses in coherent control of electron states \cite{feist2015quantum}, tailored electron phase profiles \cite{garcia2021optical}, and light-based electron aberration correctors \cite{konevcna2020electron}. Other than EEGS, special attention is also being given to time- and spatially-resolved nanothermometry.

Many recent studies have shown the feasibility of nanometer-resolved thermometry using EELS, including by means of the principle of detailed balance \cite{idrobo2018temperature, lagos2018thermometry} or by tracking the plasmon energy shift as a function of temperature \cite{chmielewski2020nanoscale}. Those experiments demonstrated the amazing possibilities for mapping  thermal effects at deep sub-optical wavelength resolution. However, they lacked time-resolution as well as the possibility to create tailored thermal gradients. With our current experimental setup, time-resolved thermometry at the nanoscale has been shown to be feasible using a diffraction-limited nanosecond-scale laser pulse combined with time-resolved monochromated EELS \cite{castioni2025nanosecond}. 

Figure \ref{Figure_PLASMON}A shows the PINEM/EELS spectra from an aluminum film during and after irradiation by a nanosecond-pulsed laser. The PINEM response and the bulk plasmon shift as a function of time can be simultaneously observed. In this particular case, PINEM curves provide the temporal profile of the laser alongside the indication of the start of the excitation, while the plasmon bulk shift provides the sample temperature given by a model \cite{castioni2025nanosecond}, as shown in Figure \ref{Figure_PLASMON}B. Finally, because this process is induced by a pulsed laser, simultaneously scanning the sample also allows us to obtain spatially and temporally-resolved maps, as shown in panel \ref{Figure_PLASMON}C obtained at the peak temperature as indicated by $t_{max}$ in panel B. One might note that with respect to previous nanothermometry experiments, the current setup allows for a controlled and time-resolved excitation source.

\section*{Perspectives}


In a recent work we have  shown that the question of the ultimate temporal resolution of Timepix3 in an electron microscope is multi-faceted \cite{auad2024time}. It includes the physical aspects of the sensor layer, such as its thickness and the applied bias voltage, and both the analog and digital parts of the application-specific integrated-circuit, i.e., the functioning of the preamplifier circuit and how the electron time-of-arrival value is determined. Specifically for electron microscopy, where fast electrons are detected, we have shown how the temporal resolution is dependent on the microscope acceleration voltage. In particular, higher energy electrons enlarge the ionization volume in the sensor layer, increasing the uncertainty of the arrival time. So, although the next generation (Timepix4) is capable of sub-200 picosecond timestamping \cite{llopart2022timepix4, heijhoff2022timing}, the prospect of reaching such values is still uncertain. Similarly to what is done in other communities, electron microscopists can seek better-suited sensor layers for fast electrons. Improvements might be made simply by reducing the sensor thickness, increasing the applied bias \cite{auad2024time} or by replacing the silicon-based sensor layer with materials such as cadmium telluride or germanium. It is worth noting that each different material presents its own advantages and challenges, and an in-depth study for fast electrons has not yet been reported in literature. Finally, it is also possible to envisage a complete technology turnaround, such as using micro-channel plates (with or without a scintillator) in order both to improve temporal resolution and reduce the detector point-spread-function \cite{tremsin2020unique}.

Recent developments in ultra-fast transmission electron microscopes \cite{feist2017ultrafast, houdellier2018development} have sparked interest in time-resolved experiments, revealing a range of quantum-optics phenomena. While many of these experiments have traditionally been conducted in ultra-fast electron microscopy, such as PINEM \cite{barwick2009photon} and electron-photonic cavity coupling \cite{wang2020coherent, henke2021integrated}, there has been increasing focus on the development of time-resolved capabilities within continuous-gun electron microscopes. One of the most promising approaches involves the use of fast electron beam blankers, which have demonstrated picosecond-range temporal resolution  \cite{verhoeven2018high, weppelman2018concept, borrelli2024direct}. These advancements underscore the critical importance of precise timing in continuous-gun electron microscopes. Time-resolved experiments hold significant promise for the future, as innovations in aberration-correction, electron monochromation, and sub-picosecond electron beams in a single microscope  could unlock new experimental possibilities.

\section*{Acknowledgement}

This project has been funded in part by the National Agency for Research under the  grant TPX4 (reference no. ANR-23-CE42-0008).
\bibliography{newbiblio.bib}

\end{document}